\documentclass[aps,pre,twocolumn,groupedaddress,showpacs]{revtex4}
\usepackage{graphicx,amsmath,amssymb}

\begin{document}
\title{On the internal modes in sine-Gordon chain}
\author{Jaroslaw E. Prilepsky}
\author{Alexander S. Kovalev}\email{kovalev@ilt.kharkov.ua}

\affiliation{B.Verkin Institute for Low Temperature Physics and
Engineering, NASU, \\ 47 Lenin Ave., Kharkov, 61103, Ukraine}

\begin{abstract}
We address the issue of internal modes of a kink of a discrete
sine-Gordon equation. The main point of the present study is to
elucidate how the antisymmetric internal mode frequency dependence
enters the quasicontinuum spectrum of nonlocalized waves. We
analyze the internal frequency dependencies as functions of both
the number of cites and discreteness parameter and explain the
origin of spectrum peculiarity which arises after the frequency
dependence of antisymmetric mode returns back to the continuous
spectrum at some nonzero value of the intersite coupling.
\end{abstract}
\date{\today} \pacs{ 05.45.-a, 63.20.Pw, 63.20.Ry} \maketitle

\thispagestyle{headings} The existence of internal modes is a well
known peculiarity of nonintegrable systems \cite{kpc98}: If one
linearizes a nonintegrable equation above the static nonlinear
solution, then the effective potential of the linear equation
obtained appears to be not reflectionless in contrast to the
integrable case. The particular example is the existence of
internal (shape) modes of kinks in the well known discrete
sine-Gordon equation (DSGE). This set of differential-difference
equations reads as (in normalized dimensionless units):
\begin{equation}\label{dsg}
\ddot{u}_n + \sin u_n + \lambda \Big[ (u_n - u_{n-1}) -  ( u_{n+1}
- u_n ) \Big]= 0 \, ,
\end{equation}
where $u_n(t)$ is the field variable, which can have a multitude
of physical meanings \cite{bk03}, $\lambda$ is a coupling
parameter, dot means the time derivative and index $n$ numerates
the 1D chain sites. The features and behaviour of kink internal
modes for DSGE, as well as for more general types of the so-called
Frenkel-Kontorova (FK) model, have been already studied in very
details \cite{bk03,bkp97,bkv90,kj00}. The spectrum of linear waves
of DSGE around a kink contains either two or one localized mode,
depending on the value of parameter $\lambda$. (We consider the
stable kink centered between the chain sites.) The frequencies of
these modes lie inside the spectrum gap; the linear spectrum
itself is given by
\begin{equation}\label{spec}
\omega^2 = 1 + 4 \lambda \sin^2 \frac{k}{2} \, ,
\end{equation}
where $\omega$ is the frequency of linear waves and $k$ is the
wave number. The lowest (gap edge) frequency is $\omega =1$ (in
the renormalized units).

However, up to our knowledge, notwithstanding the great quantity
of results none so far has concentrated on the way an internal
modes ceases to exist: Only the fact of existence or nonexistence
of that mode has been the subject of interest
\cite{kpc98,bk03,bkp97,bkv90,kj00}. In this paper we focus on the
effect entailed by the detachment of the internal mode frequency
from the spectrum. So, the question arises: How the internal
modes, which split off from the spectrum at some nonzero value of
coupling parameter (or some other effective parameter altering the
system state), affect the spectrum itself and how they behave
while still being inside the spectrum.

To begin with, we note that the analytical approximate solution
for a static kink of DSGE (\ref{dsg}) can be found either in the
so-called anticontinuum limit (extremely small values of
$\lambda$) as a series in powers of $\lambda$ \cite{bkv90}, or in
the opposite case of strong coupling (large $\lambda$), when the
discrete kink solution acquires the form of that in the continuous
SGE with small corrections due to discreteness (see e.g.
\cite{fk96}). Let us now summarize the results on the existence of
internal modes in the DSGE containing a single kink. For small
$\lambda$ only the lowest symmetric Pierls-Nabarro (PN) mode,
associated with the kink oscillations in the PN relief, has the
frequency inside the gap: it corresponds to the translational kink
mode of continuous SGE, activated due to discreteness. This mode
remains the internal mode for the whole interval of kink
stability. The properties of this mode are well understood and
studied \cite{bk03,bkv90}, and we shall be mainly concerned with
the other internal mode. For larger (but still weak) $\lambda$
there exists the "critical" point $\lambda_d$, where one more mode
dependence detaches from the continuous spectrum: this mode
corresponds to the oscillation of kink width. (More complicated
on-site FK-type potentials may have a larger variety of localized
internal eigenmodes \cite{bkp97} or, for the case of more than one
kink in the chain, the DSGE possesses a larger number of internal
modes as well \cite{bk03}.) The existence of the second internal
mode for large $\lambda$ obviously matches the criterium given by
Kivshar et al. in Ref.\cite{kpc98} for nearly integrable SGE (see
also Ref.\cite{kj00} for details).

\begin{figure}\centering
\scalebox{0.5}[0.5]{ \includegraphics[bb=63 324 511 795 ]{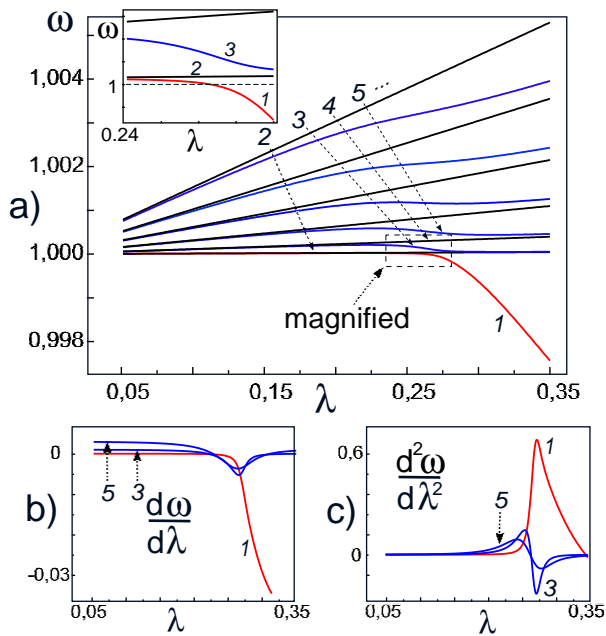}}
\caption{\label{f1} Numerical results for DSGE containing 250
sites with a kink centered in the middle of the chain between the
sites. a) The frequency dependencies for several lowest modes on
the value of $\lambda$ (the dependence for the lowest PN-mode is
not presented); the numbers correspond to the mode number, i.e. to
the number of nodes of corresponding eigenfunction. The curve
marked { \em ''1''} is the dependence for localized antisymmetric
mode. The inset shows the magnified region in the close vicinity
of the detachment value $\lambda_d$ (the bandgap threshold marked
by the dashed line) . The bottom panels show the behaviour of b)
$\mathrm{d} \omega / \mathrm{d} \lambda$ and c) $\mathrm{d}^2
\omega / \mathrm{d} \lambda^2$ for three lowest antisymmetric
modes ({\em 1,3} and {\em 5}) . }
\end{figure}

The effects we shall discuss can be readily seen in Fig.\ref{f1}:
As the internal mode frequency goes back for small $\lambda$ to
the quasicontinuous spectrum, it brings about a conspicuous change
in the behaviour of higher modes. However only the dependencies of
odd modes (the mode number corresponds to the number of
eigenfunction nodes) ''feel'' the return of the localized first
antisymmetric mode to the spectrum. The modes with the even number
of eigenfunction nodes do not react to this return. So, we can
infer that only the modes having the same symmetry as the mode,
which detaches from the spectrum, experience pronounced change
(which can be also seen in the behaviour of the derivatives,
Fig.\ref{f1}b,c). This change ''propagates'' inwards the spectrum
becoming less and less abrupt with the increase of mode number. In
addition, as can be seen in the inset of Fig.\ref{f1}a, the
dependence of the first antisymmetric mode indeed enters the
spectrum: it crosses the band edge frequency $\omega=1$ at the
point $\lambda_d \approx 0.26$ (of course, it crosses only the
band threshold but not any of the frequency dependencies). For
smaller $\lambda$ the dependence of former antisymmetric localized
mode belongs to the spectrum of usual delocalized waves, tending
to coalesce with the dependence of second (symmetric) mode.
Also we note another interesting feature of eigenfrequency
dependencies. For large $\lambda$, i.e. when we are approaching
the continuum limit, the dependencies for odd modes with numbers
$2n+1$ tend to coalesce with the dependencies for {\em lower} even
modes, i.e. with those having numbers $2n$. However the detachment
of one mode results in another tendency: after the splitting, for
$\lambda$ close to zero, the dependencies for odd modes tend to
coalesce with those of { \em higher} even modes, i.e. the
frequencies of modes with numbers $2 n +1$ come closer and closer
to those for the modes with numbers $2n+2$. The inflection of the
odd modes dependencies and peculiarities of their derivatives in
the vicinity of $\lambda_d$ merely indicate this ''change'' in the
tendency for coalescence.

To delineate and compare the behaviour of internal modes we first
study some illustrative examples when we have only a few pendulums
with linear bonds.

{\em Small number of pendulums with linear bonds.} For the
simplest case of two pendulums the dynamical equations read as:
\begin{equation}\label{2p}
\ddot{u}_{1, \, 2}+\lambda (u_{1, \, 2}-u_{2, \, 1}) + \sin u_{1,
\, 2} = 0 \, .
\end{equation}
Linearizing Eqs.(\ref{2p}) around the ground state solution, $u_1
= u_2 = 0$, we arrive at the eigenfrequency dependencies:
$\omega_0^2=1$ , $\omega_1^2 = 1+2 \lambda$. The higher frequency
corresponds to the antiphase oscillations, the lower one -- to the
inphase motion. Now let us study the spectrum of linear
oscillations above the inhomogeneous static solution of
Eq.(\ref{2p}) (analogue of a kink). First we have to find the
static distribution for small values of $\lambda$. We suppose that
this distribution has a form: $u_1=u^0$, $u_2 = 2 \pi - u^0$,
$u^0=\alpha \lambda + \beta \lambda^2 + \mathcal{O}(\lambda^3)$,
and we should determine the corresponding constants, $\alpha$ and
$\beta$, in the dependence $u^0(\lambda)$. Expanding the sine in
Taylor series and equalizing coefficients at the same powers of
$\lambda$ we find:
\begin{equation}\label{2u}
u^0 = 2 \pi \lambda - 4 \pi \lambda^2 + \ldots \, .
\end{equation}
Then we linearize the initial set of dynamical equations
(\ref{2p}) with respect to small quantities $v_i$: $u_1 = u^0 +
v_1 \mathrm{e}^{\mathrm{i} \omega t}$, $u_2= 2 \pi - u^0 + v_2
\mathrm{e}^{\mathrm{i} \omega t}$. The consistency condition gives
the spectral dependencies: for the lower inphase mode $\omega_0^2
= \cos u^0$, that is the analogue of PN mode in a large system,
and for the antiphase mode $\omega_1^2 = \cos u^0 + 2 \lambda$ ---
the analogue of the second antisymmetric internal mode. The
analytical form for the frequencies, which is valid for the small
$\lambda$ values, is as follows:
\begin{equation}\label{2spec}
\omega_0^2 = 1 -2 \pi^2 \lambda^2 + \ldots \, , \qquad \omega^2_1
= \omega_0^2 + 2 \lambda \, .
\end{equation}
The frequency dependence $\omega_1(\lambda)$ grows linearly with
$\lambda$ for weak coupling parameter values and stays inside the
band, but then, as $\lambda$ becomes larger, this growth slows
down showing an ultimate tendency of this mode to get inside the
gap below. The value of $\lambda$ for which this dependence
crosses the band threshold can be roughly estimated as $\lambda_d
\approx 1/ \pi^2$. One also notes that the frequency of PN-mode,
$\omega_0$, drops to zero for $u^0 = \pi/2$. This manifests the
well known fact that for the finite size chain there occurs an
instability when the characteristic spatial scale of a kink
overgrows the size of a system (see e.g. Ref.\cite{k78,kp03}). For
a large system the PN-mode softens at some critical value
$\lambda_c \sim L^2$, where $L\gg1$ is the system size
\cite{kp03}. For two pendulums, setting $u^0_1 = \pi/2 - U$,
$u^0_2 = 3 \pi/2 + U$, $U\ll1$, one finds the frequency
dependencies near the critical point: $\omega_0^2 \approx
(\lambda_c - \lambda)/2 \lambda_c^2$, with $\lambda_c = 1/\pi$,
and $\omega_1^2 = \omega_0^2 + 2 \lambda$. We see that at the
critical point $\omega_1^2(\lambda_c) = 2/ \pi <1$, i.e. the
frequency remains inside the gap. Thus we expect that this
internal mode dependence for a big system does not return back to
the spectrum for large values of $\lambda$ and continues to be the
internal mode for the whole interval of kink stability except for
the extremely weak coupling. The splitting of this mode dependence
from the continuum spectrum is of the order of $\lambda^{-1}$ for
large values of coupling parameter \cite{kpc98,kj00}.

The systems containing more sites enable us to study the '''degree
of localization'' since we now have the outer and inner (core)
sites. Let us mark the sites in the kink core (the middle of a
chain) by numbers ''1'' and ''-1'', their neighbor sites by ''2''
and ''-2'' etc. For the case of four pendulums the ''kink''
solution for small $\lambda$ is given by:
\begin{equation}\label{4u}
u_2^0 \approx 2 \pi \lambda^2 + \ldots\, ,\qquad u_1^0 \approx 2
\pi \lambda - 8 \pi \lambda^2 + \ldots \, ,
\end{equation}
and the obvious solution symmetry, $u_{-i} = 2 \pi - u_{i}$, gives
the field distribution for the remaining sites. In this limit the
spectral dependencies for two lowest modes
are:
\begin{subequations}
\begin{gather}\label{4pend-om0}
\omega_0^2 \approx 1 - \pi^2 \lambda^2 \, , \\ \omega_1^2 \approx
1+(2 - \sqrt 2) \lambda - \frac{\pi^2}{ 2}( 2 - \sqrt 2) \lambda^2
\, .\label{4pend-om1}
\end{gather}
\end{subequations}
Putting $\omega_1(\lambda)=1$ in Eq.(\ref{4pend-om1}) one gets the
rough estimate for the point where this frequency dependence
enters the ''quasicontinuum'' spectrum: $\lambda_d \approx 2/
\pi^2$. In the same way one could deal with the systems of bigger
size (see below).

\begin{figure}
\centering
\scalebox{0.55}[0.55]{\includegraphics[65,462][509,792]{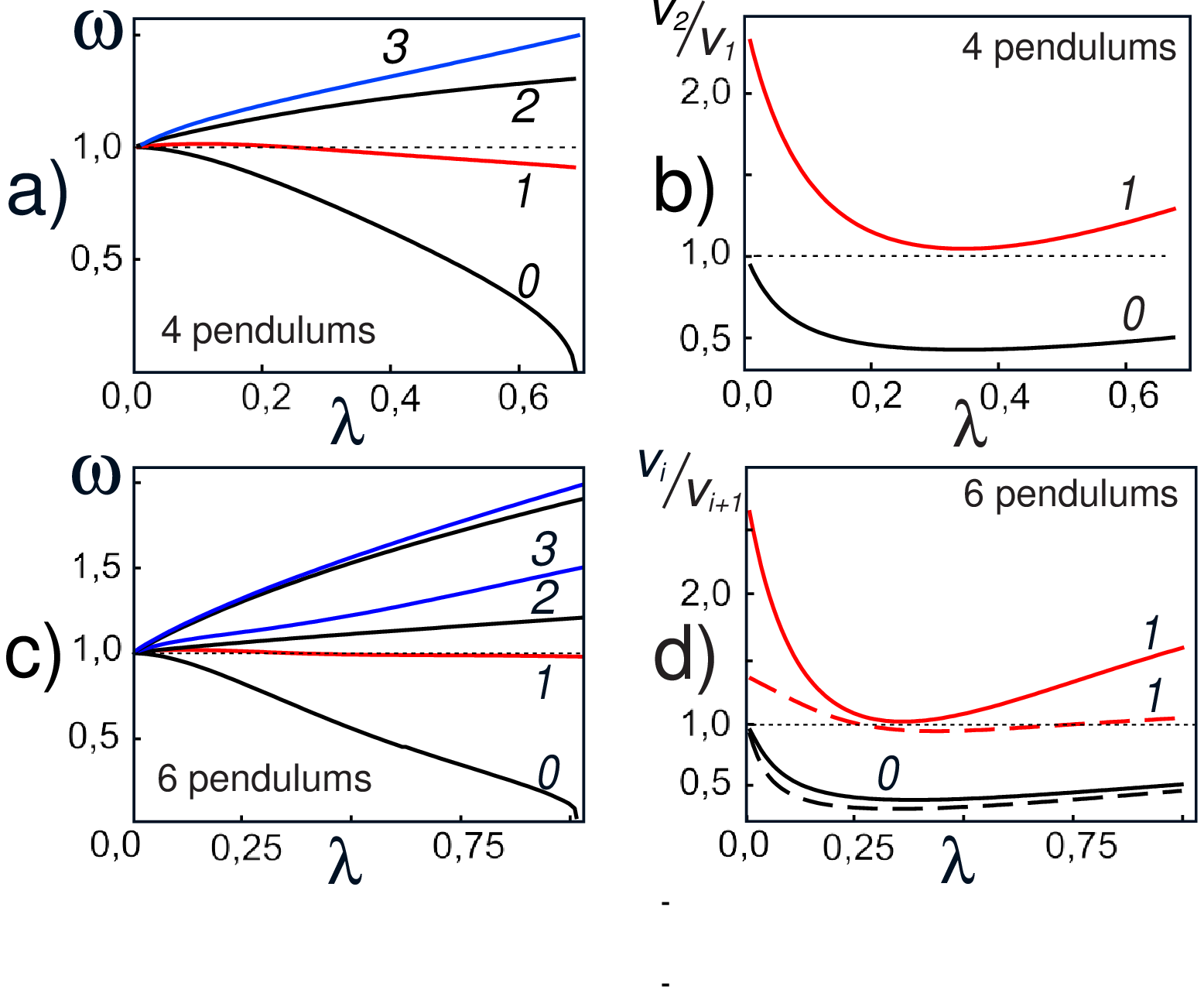}}
\caption{Numerical results for DSGE containing four and six sites
with a ''kink'' centered in the middle between the sites. Left
panels: full spectrum of a) 4 pendulums and c) 6 pendulums in the
interval of kink stability; the numbers correspond to the mode
number. Right panels show the dependencies of oscillations
amplitude ratios of different neighbor sites corresponding to two
lowest modes: b) for four pendulums $v_2/v_1$; d) for six
pendulums $v_{i+1} / v_{i}$ with $i=1$ (solid line) and $i=2$
(dashed line). The digits correspond to the mode number.}
\label{f2}
\end{figure}

The spectrums and amplitude ratio dependencies on $\lambda$ for
four and six pendulums are given in Fig.\ref{f2}. As the number of
sites grows, the spectrum peculiarities, seen in Fig.\ref{f1},
become more and more pronounced. In the panels (b) and (d) of
Fig.\ref{f2} we present the dependencies for the ratios of the
oscillation amplitudes for different pairs of neighbor sites,
$v_{i+1}/v_i$, for PN (zeroth) mode and first antisymmetric mode.
Both modes display the tendency for localization in the kink core
with the growth of $\lambda$, but coming closer to the critical
point $\lambda_c$ they become localized worse because of the large
kink width.
Compared with the first antisymmetric mode, the PN-mode appears to
be localized better. These dependencies have a minima for some
intermediate value of coupling, where the inhomogeneity brought by
the kink is pronounced well enough to sustain the modes
localization and, from the other side, the kink wight  is still
not large.

\begin{figure}
\centering
\scalebox{0.6}[0.6]{\includegraphics[90,461][469,772]{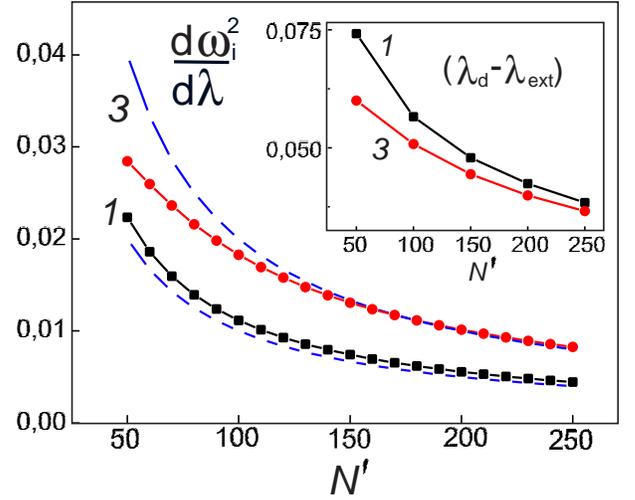}}
\caption{The dependencies for the modulus of derivative at the
point $\lambda = \lambda_d$ on the number of sites for the first
(antisymmetric) mode and for the next, third mode. The digits
correspond to the mode number. Dashed curves for each mode show
the analytical prediction obtained by virtue of Eqs.(\ref{om1-3}).
The inset shows the dependence of deviation $\lambda_d -
\lambda_{\mathrm{ext}}$ of the extremal point from the detachment
point as a function of number of sites $N' = 2 N$.} \label{f3}
\end{figure}

{\em Large systems.} Now suppose that our system has $2N$ sites.
The static distribution for the kink is again given by $u^0_1 = 2
\pi \lambda + \mathcal{O}(\lambda^2)$, $u^0_n = \mathcal{O}
(\lambda^n)$. Then we substitute $u_n(t) = u^0_n + v_n
\mathrm{e}^{\mathrm i \omega t}$, and linearize the system
obtained with respect to $v_n$ noting that for the antisymmetric
modes the symmetry is $v_n = - v_{-n}$. After that one gains the
following linear system:
\begin{equation}\label{ant}
\begin{split}
\begin{cases}
&(\omega^2 - 3 \lambda - \cos u^0_1) v_1 + \lambda v_2 =0
\, , \\
& \; \vdots \\
&(\omega^2 - 2 \lambda - \cos u^0_n) v_n + \lambda (v_{n-1}+v_{n+1}) =0 \, , \\
& \; \vdots \\
&(\omega^2 - \lambda - \cos u^0_N) v_N + \lambda v_{N-1} =0 \, .
\end{cases}
\end{split}
\end{equation}
For the symmetric modes ($v_n = v_{-n}$) the first equation of
system (\ref{ant}) is to be replaced with
\begin{equation}\label{sym}
(\omega^2 - \lambda - \cos u^0_1) v_1 + \lambda v_2 =0 \, .
\end{equation}
For small $\lambda$ we use the approximate expressions for
$u_n^0$, and in the leading approximation substitute $\cos u_1^0 \
\to 1 - 2 \pi^2 \lambda^2$, $\cos u_n^0 \to 1$ for $n>1$. This
means that we effectively replaced the kink with two isotopic
impurities located at sites {\em 1} and {\em -1}, the ''strength''
of which alters with the change of $\lambda$. Seeking the solution
in the form $v_n = A \mathrm{e}^{\kappa n} + B \mathrm{e}^{-
\kappa n}$, with constant $A$ and $B$, one obtains the spectral
dependence for $\omega(\kappa)$ in form (\ref{spec}), where
$\kappa = \mathrm{i} k$. Then using the consistency condition
after some straightforward algebra we arrive at the relation which
defines the allowed values of $\kappa$ for antisymmetric modes:
\begin{eqnarray}\nonumber
(1 &-& 2 \pi^2 \lambda) \sinh [\kappa (N-1)] \\ + 2 \pi^2
&\lambda& \sinh [\kappa N] - \sinh[\kappa(1+N)] = 0 \, .
\label{rel1}
\end{eqnarray}
Expanding this relation in the vicinity of $\kappa =0$ ($\kappa N
\ll 1 $) up to $\kappa^5$ we determine the allowed values of
$\kappa$ for the frequency of first antisymmetric mode,
$\omega_1$, and for the next, third mode, $\omega_3$, from the
biquadratic equation: $a \kappa^4 + b \kappa^2 + c =0$. The
expressions for the coefficients are:
\begin{gather}\label{coefs} \nonumber
a = \frac{1  (\lambda - \lambda_d) + 5(2 N^2 +
N^4)(\lambda - \lambda_d) - 5 \lambda ( 2 N^3 + N)}{60} \, , \\
\nonumber b = (\lambda - \lambda_d)/3 + N^2 (\lambda - \lambda_d)
- N \lambda \,, \\ \nonumber c = 2 (\lambda-\lambda_d) \, ,
\end{gather}
and the detachment point is $\lambda_d = 1/\pi^2$. The notable
fact, which can be extracted from Eq.(\ref{rel1}), is the
following: if $\lambda$ is close to zero, $\lambda \ll N^{-1}$, we
have $\kappa^2 = - k^2 \sim N^{-2}$, and for this values of
coupling parameter one finds the expressions for frequency
dependencies as:
\begin{subequations}\label{om00}
\begin{gather}\label{om0}
\omega_1^2 \approx 1 + 2 \lambda (3- \sqrt 3) N^{-2} \, , \\
\omega_3^2 \approx 1 + 2 \lambda (3+ \sqrt 3 ) N^{-2}  \, .
\end{gather}
\end{subequations}
So, initially, for small $\lambda$, the lowest antisymmetric mode,
Eq.(\ref{om0}), goes up being inside the spectrum. However quite a
different situation occurs if one moves inside the region where
the inequality $|\lambda-\lambda_d| \ll N^{-1}$ holds, i.e. in the
close vicinity of $\lambda_d$. In this region one obtains
$\kappa^2 \sim N^{-1}$ (here the inequality $\kappa N \ll 1$ is
true because of the additional smallness provided by the factor
$(\lambda - \lambda_d)$). Then for the frequency dependencies we
have:
\begin{subequations}\label{om1-3}
\begin{gather}\label{om1}
\omega_1^2  \approx 1 - 2  (\lambda - \lambda_d)N^{-1} \, , \\
\label{om3} \omega_3^2 \approx 1 - 4(\lambda - \lambda_d) N^{-1}+
6 \lambda_d N^{-2} \, .
\end{gather}
\end{subequations}
The second derivative of $\omega_1$ at the point $\lambda =
\lambda_d$ involves the term independent on $N$: $\mathrm{d^2}
\omega_1 / \mathrm{d} \lambda^2 = - \lambda_d^{-1}$. Therefore we
can conclude that in the limit $N \to \infty$ the splitting of
this dependence from the lower boundary of the spectrum must have
a parabolic form. From the expressions (\ref{om00},\ref{om1-3}) it
becomes evident what brings about the peculiarity as $\lambda$
approaches $\lambda_d$, i.e. as the first antisymmetric mode
detaches from the spectrum. For $\lambda$ close to zero we have
$\mathrm{d} \omega_{1,3} / \mathrm{d} \lambda \sim N^{-2}$,
whereas in the vicinity of $\lambda_d$ the different dependence
takes place: $\mathrm{d} \omega_{1,3} / \mathrm{d} \lambda \sim
N^{-1}$. (Note that the sign of the first derivative also
changes.) Because of this the initial weak ($\sim N^{-2}$) growth
for small values of $\lambda$ changes to more rapid ($\sim
N^{-1}$) decrease in the close vicinity of $\lambda_d$. The
dependencies for the modulus of first derivatives at $\lambda =
\lambda_d$ on the number of sites shown in Fig.\ref{f3}. For
$N'=2N \gtrsim 30$ they are in a good agreement with analytical
results (\ref{om1-3}). With the increase of $N$ this agreement
becomes better inasmuch as we omitted the higher with respect to $
N^{-1}$ terms in Eqs.(\ref{om1-3}). In the region $\lambda <
\lambda_d$ there must be an extremum point
$\lambda_{\mathrm{ext}}$, $\mathrm{d} \omega_i / \mathrm{d}
\lambda |_{\lambda_{\mathrm{ext}}} = 0$, for both dependencies
$\omega_{1,3}(\lambda)$, at which the monotonic growth changes to
decreasing. These extremum points, $\lambda_{\mathrm{ext}}$, tend
to $\lambda_d$ as $N$ gets bigger: $(\lambda_d -
\lambda_{\mathrm{ext}}) \sim N^{-1}$. The dependencies of the
value of difference, $(\lambda_d - \lambda_{\mathrm{ext}})$, on
the number of sites $N$ are presented in the inset panel of
Fig.\ref{f3}. For $\lambda>\lambda_d$ the first antisymmetric mode
gets into the spectrum gap and becomes an internal mode. However
the dependence for the next, third antisymmetric mode, in spite of
its tendency to drop down, cannot cross the dependence of the
preceding second (symmetric) mode and therefore these two
dependencies get closer and closer to each other.

Now consider an infinite system. We seek the solution in the form
of localized wave, $v_n \sim \mathrm{e}^{-\kappa n}$, $\kappa>0$.
Whereupon the only condition defining the allowed values of
$\kappa$ becomes as follows:
\begin{equation}
\mathrm{e}^{- \kappa} + 2 \pi^2 \lambda - 3 - 4 \lambda \sinh^2
\frac{\kappa}{2} = 0 \, .
\end{equation}
Expanding this relation we arrive at the dependence: $\kappa_1 = 2
\lambda_d^{-1} (\lambda - \lambda_d)$, and then one finds the
expression for the frequency of the antisymmetric localized mode
as:
\begin{equation}\label{om1inf}
\begin{cases}
\omega_1^2 = 1 \, , \quad &\text{for} \; \; \lambda < \lambda_d \, ,\\
\omega_1^2 \approx 1 - 4 \lambda_d^{-1} (\lambda - \lambda_d)^2 ,
\quad &\text{for} \; \; \lambda> \lambda_d \, ,
\end{cases}
\end{equation}
in consistency with the result for finite system. The
characteristic spatial scale for the eigenfunction of this
internal mode is $l_1 = \kappa_1^{-1}$. Obviously, the smaller
$l_1$, the better the eigenfunction is localized. We see that at
the outset, as this mode detaches, $l_1 \sim (\lambda -
\lambda_d)^{-1}$.

Proceed to studying the symmetric modes. The relation which
defines the allowed values of $\kappa$ for those reads:
\begin{eqnarray}\nonumber
(1 &+& 2 \pi^2 \lambda) \sinh [\kappa (N-1)] \\ - 2 (1 + \pi^2
&\lambda&) \sinh [\kappa N] + \sinh[\kappa(1+N)] = 0 \, .
\label{symk}
\end{eqnarray}
Then we again expand this relation for $\kappa N \ll 1$ up to the
fifth power and determine $\kappa$ from the biquadratic equation:
$d \kappa^4 + g \kappa^2 + f =0$, where
\begin{gather}\nonumber
d = \frac{ 5 N (\lambda + \lambda_d) + 10 N^3 (\lambda + \lambda_d
) -\lambda(5 N^4 + 10 N^2 +1)}{60} \, , \\ \nonumber f = N
(\lambda + \lambda_d) - (N^2 + 1/3) \lambda \,, \\ \nonumber g =
-2 \lambda \, .
\end{gather}
In the region $\lambda \ll N^{-1}$ one obtains:
\begin{equation}\label{om+0}
\omega_0^2 \approx 1 - 2 \lambda_d^{-1} \lambda N^{-1}  \, ,
\qquad \omega_2^2 \approx 1 + 6 \lambda N^{-2} \, .
\end{equation}
The dependencies for symmetric modes have neither an inflection
point nor any peculiarity at $\lambda = \lambda_d$.

For the infinite system the relation for $\kappa$ is written as
follows:
\begin{equation}
\mathrm{e}^{- \kappa} + 2 \pi^2 \lambda - 1 - 4 \lambda \sinh^2
\frac{\kappa}{2} = 0 \, .
\end{equation}
Then we arrive at the dependence for PN-mode for weak coupling in
the form:
\begin{equation}\label{om0inf}
\omega_0^2 \approx 1 - 4 \lambda_d^{-2} \lambda^3 \, ,
\end{equation}
which is in agreement with this dependence given in
Ref.\cite{bkv90}. The localization distance now is: $l_0 =
\kappa_0^{-1} \sim \lambda^{-3/2}$. Thus we can infer that the
localization for PN-mode with the growth of coupling value
$\lambda$ develops faster than for antisymmetric internal mode, as
it was pointed above while considering small size systems (cf.
Fig.\ref{f2}, right panels).

The results for the frequency dependencies
(\ref{om1inf},\ref{om0inf}) could also be obtained with the use of
Green function (Lifshitz) technique for a linear chain with
impurities, (see the monograph \cite{k99}), which was employed in
Refs.\cite{bkp97,bkv90}. However, for the sake of getting simple
analytical expressions one should replace the kink solution with
the finite number of effective impurities. In the case of two
impurities model, utilized in the current paper, one would obtain
{\em exactly the same results}. In fact, the usage of the two
impurities model means that we retain only leading terms in powers
of parameter $\lambda$.

To sum up, we investigated the features of spectrum and kink
internal modes of DSGE focusing on the effect of detachment of an
internal mode  from the quasicontinuous spectrum. We note that the
two-impurities model, used for our analysis, gives a somewhat
different value for the detachment point ($\lambda_d = \pi^{-2}$,
while the numerically estimated value for DSGE is $\lambda_d
\approx 0.26$). This is explained by the fact that the value of
$\lambda_d$ is nonzero, whereas we had to use the asymptotical
expansions valid for $\lambda \to 0$. This means the higher order
terms with respect to $\lambda$ should always result in the
corrections to the analytically obtained value of $\lambda_d$.
However the two impurities model allowed us to obtain analytical
results describing the spectrum peculiarities, which are in good
agreement (qualitative and even quantitative for large $N$) with
the numerical ones.  For the infinite system the dependence of
first antisymmetric mode detaches smoothly (with zero first
derivative) from the band edge, but for the finite system it does
cross the bandgap threshold. For extremely small values of
$\lambda$ the first mode belongs to the spectrum. The initial weak
growth ($\sim N^{-2}$) of lowest antisymmetric modes frequencies
with the value of $\lambda$ then changes to more rapid ($\sim
N^{-1}$) decreasing. The extremum point $\lambda_{\mathrm{ext}}$
approaches the detachment point $\lambda_d$ as the number of sites
gets bigger. At $\lambda = \lambda_d$ the first antisymmetric mode
drops into the spectrum gap. The higher antisymmetric mode come
closer and closer to the preceding symmetric modes with the
increase of coupling but this tendency weakens with the growth of
mode number. The symmetric modes do not display any peculiarity at
$\lambda_d$.

\end{document}